\documentstyle[preprint,aps]{revtex}
\draft
\preprint{\vbox{\baselineskip=14pt
\rightline{hep-ph 9512255}\break
\rightline{OCIP/C-95-14}\break
\rightline{December 1995}}}

\begin{document}

\title{\bf Photon and $Z$ Induced Heavy Charged Lepton Pair Production
at a Hadron Supercollider}
\author{G. Bhattacharya, Pat Kalyniak and K. A. Peterson}
\address{Ottawa-Carleton Institute for Physics\\
Department of Physics, Carleton University\\
1125 Colonel By Drive,
Ottawa, Ontario, Canada K1S 5B6}

\maketitle

\begin{abstract}
{We investigate the pair production of charged heavy
leptons via photon-induced processes at the proposed CERN Large Hadron
Collider (LHC). Using effective photon and $Z$
approximations, rates are given for $L^+L^-$ production due to $\gamma \gamma$
fusion and $Z \gamma$ fusion for the cases of inelastic, elastic
and semi-elastic $pp$ collisions. These are compared with the corresponding
rates for production via the gluon fusion and Drell-Yan mechanisms. Various
$\gamma \gamma$ and $Z \gamma$ differential luminosities for $pp$ collisions
are also presented.}    \end{abstract}


\newpage
\section{Introduction}

There has been a recent revival of interest in processes mediated via photons
at $e^+e^-$ colliders\cite{eecol}. In the present work we address the question
of photon induced production processes at hadron colliders, which
have been relatively neglected. Since photons have a universal coupling to
charged fermions it seems natural to consider fermion pair production as a
benchmark process.

Heavy charged leptons are a feature of many models which extend the particle
content of the Standard Model (SM). These include models which propose a
complete additional generation\cite{gen} of heavy quarks and leptons, as well
as those like $E_6$ based superstring-inspired models\cite{e6}, which contain
extra particles within each generation. The Drell-Yan\cite{dy}, gluon fusion
\cite{gf}, and weak gauge boson fusion\cite{gbf} mechanisms for the production
of heavy charged leptons in hadron collisions have been investigated in the
past.
We here consider heavy charged lepton pair production in hadronic collisions
via
two photon ($\gamma \gamma$) and $Z \gamma$ fusion. In Fig. 1, we show the
Feynman diagrams for the relevant subprocesses. We assume standard model
couplings for both the $\gamma$ and the $Z$ boson to the heavy leptons, $L$.

The inelastic process $pp \rightarrow \gamma \gamma X \rightarrow L^+  L^-X$
(where the photons come from the quarks), the elastic process $p p \rightarrow
\gamma \gamma p p \rightarrow L^+L^- p p$ (where the photons come from the
protons),   and the semi-elastic process $pp \rightarrow \gamma \gamma p X
\rightarrow L^+L^- p X$, are considered in a
Weizs\"{a}cker-Williams approximation\cite{ww,elww} (WWA). We
have previously   presented some results for these photon fusion
processes\cite{us}, as have some other authors \cite{do}. The inelastic $\gamma
\gamma$ process was considered on its own by Eboli {\sl et al.}\cite{eboli}
also.

Here, we also present heavy charged lepton pair production via $Z \gamma$
fusion. In fact, the $Z \gamma$ fusion process yields a larger production cross
section than the $\gamma \gamma$ fusion over most of the lepton mass range we
consider ($m_L$  of 100 - 700 $\rm GeV$). Our $Z \gamma$ fusion calculation is
done within the framework of an effective vector boson
approximation\cite{kane,dawson}, for both inelastic and semi-elastic $pp$
collisions. Heavy charged lepton production via  $Z\gamma$ fusion has been
previously considered for the case of inelastic $pp$ collisions by Eboli {\sl
et al.}\cite{eboli} We do not agree with their results and discuss the origin
of that discrepancy.

The organization of this paper is as follows.
We describe in Sec. 2 the details of the effective photon and effective $Z$
approximations which we have used. The $V \gamma$ ($V = \gamma, Z$)
luminosities in $pp$ collisions are presented in Sec. 3; these results are of
broader interest than just for the particular heavy lepton production
considered here. The $L^+L^-$           production cross sections are
presented in Sec. 4 and we summarize in the final section.

\section{Effective Photon and Effective $Z$ Approximations}

The central idea of the effective photon or
Weizs\"{a}cker-Williams approximation\cite{ww} is to approximate the
scattering cross-section involving a charged particle by a convolution of the
equivalent number of photons in that particle with the relevant photoproduction
cross-section. The quark-parton model has been very successful in describing
hadronic interactions at high energies. For inelastic $pp$ scattering a proton
may be regarded essentially as a collection of freely travelling elementary
constituent quarks and gluons.
The Weizs\"{a}cker-Williams
photon spectrum, $f_{\gamma / q}(x)$, from a quark (of charge $e_q$) is
given by\cite{brodsky}
\begin{equation}
f_{\gamma / q} (x) = \frac{e_{q}^2}{8 \pi^2} \frac{1 + (1 - x)^2}{x}
\log \frac{t_{\rm {max}}}{t_{\rm {min}}}.
\end{equation}

\noindent
where $x$ is the fraction of the quark energy carried off by the photon.
Here $t_{\rm {max}}$ and $t_{\rm {min}}$ are the characteristic maximum and
minimum photon momentum transfers. For the process under consideration, pair
production of heavy charged leptons of mass $m_L$, we have taken these to be
$t_{\rm {max}} = \hat{s} /4 - m_{L}^2$, and $t_{\rm {min}} = 1$ ${\rm GeV}^2$,
with $\hat{s}$ being
the center of mass energy in the parton frame. There is some flexibility in
the choice of $t_{\rm {max}}$. However, in agreement with
Altarelli {\sl et. al.}\cite{ammr}, we have found that our results are not
very sensitive to this
parameter, within the limits of the Weizs\"{a}cker-Williams approximation.
The particular choice of the minimum momentum transfer, $t_{\rm {min}}$,
guarantees that the photons are obtained from the deep inelastic scattering
of protons, when the quark-parton model is valid.

For inelastic scattering the incident proton ceases to exist and the
constituents hadronize. However for elastic $pp$ scattering the incident
protons remain intact. The photon
spectrum from protons for the elastic case, $f^{\rm{el}}_{\gamma / p}$, has
been derived by Kniehl\cite{elww} in a modified Weizs\"{a}cker-Williams
approximation. It takes the form given below.

\begin{equation}
f^{\rm{el}}_{\gamma / p} (x) = -\frac{\alpha}{2 \pi} x \int_{-\infty}^{t_{\rm
{max}}}\frac{dt}{t} \left \{2 \left[ \frac{1}{x} \left( \frac{1}{x} - 1 \right)
+ \frac{m_p^2}{t} \right] H_1(t) + H_2(t) \right \}
\end{equation}

\noindent Here, $t_{\rm {max}} = -m_p^2 x^2/(1 - x)$ while $H_1$ and $H_2$
are functions of the
electric and magnetic form factors of the proton of mass $m_p$, as given below.

\begin{equation}
H_1(t) = \frac{G_E^2(t) -(t/4m_p^2)G_M^2(t)}{1 - t/4m_p^2} , \; \;\; \;
H_2(t) = G_M^2(t)
\end{equation}

\noindent The form factors are parametrized as

\begin{equation}
G_E(t) = (1 - t/0.71 \, {\rm GeV}^2)^{-2} , \; \;\; \; G_M(t) = 2.79 G_E(t).
\end{equation}
Using these parametrizations Kniehl has obtained a closed analytic
expression\cite{elww} for $f^{\rm{el}}_{\gamma /p}$, which we have used in our
calculation.

For hadron interactions at very high energies in the multi-TeV range, one can
consider the $W$ and $Z$ bosons also as constituents of protons. In analogy
with
the effective photon approximation, effective vector boson approximations have
been  developed which considerably simplify calculations involving gauge
bosons.
However unlike photons, a massive gauge boson has both transverse and
longitudinal polarizations and these degrees of freedom must be treated
separately. The effective $W$ and $Z$ approximations give the spectra of
longitudinally and transversely polarized $W$'s and $Z$'s from quarks in the
protons.

We have done our calculations using two different forms for the effective
$Z$ approximation, the non-leading order Z distributions of Dawson
\cite{dawson}
and those of
Johnson {\sl et. al.}\cite{johnson}. In
the original formulation of the approximation, Dawson      \cite{dawson}
presents the distribution function of $Z$ bosons in quarks both  in the leading
logarithmic (LL) approximation and also gives non-leading        order
corrections up to ${\cal O} (M_Z^2/E_q^2)$, where $E_q$ is the quark energy. We
used the non-leading order distribution in our calculation. An improved
formulation of the effective vector boson approximation has been presented
by Johnson {\sl et. al.}\cite{johnson}. This formulation does not invoke any
kinematic approximations made in the original formulation, and
claims to improve the accuracy and extend the range of applicability of the
effective vector boson approximation\cite{johnson,godbole}. Dawson
presents the distribution of transverse Z bosons averaged over the two
transverse polarizations whereas Johnson {\sl et. al.} derive
separately each of the positive and negative helicity polarizations.
The distribution expressions are too lengthy to reproduce here
\cite{dawson,johnson,godbole}.

Johnson {\sl et. al.}, in comparing their vector boson distribution functions
with that of the leading logarithmic (LL) distribution functions of Dawson,
find the LL expressions to be a considerable overestimate, particularly for
small $x$ and for the transverse polarization. We here find rather good
agreement between our $Z\gamma$ luminosity results using the distribution
functions of Dawson including non-leading order
corrections and those of Johnson {\sl et. al.}, as we show in the next
Section.

\section{Photon-Photon and $Z$-Photon Luminosities}
Using the effective photon and $Z$ approximations described above we
present here the two photon and $Z$-photon differential luminosities in
$pp$ collisions as a function of
$\tau$ (the ratio of the $\gamma\gamma/Z\gamma$ subprocess energy and the total
$pp$ energy), for the inelastic, elastic and semi-elastic cases. These results
are useful for adaptation to the production of other particles than heavy
leptons.

The $V \gamma$ $(V = \gamma, Z)$ differential luminosity for inelastic $pp$
collisions is given by,

\begin{equation}
\Bigl.\frac{dL^{\rm{inel}}}{d\tau}\Bigr|_{V \gamma/pp} = \sum_{i,j} \int
_{\tau}^{1} \frac{d\tau^\prime}{\tau^\prime} \int_{\tau^\prime}^{1}
\frac{dx}{x} f_{q_{i}/p}(x) f_{q_{j}/p} (\tau^\prime/x) \Bigl.
\frac{dL}{d\xi}\Bigr|_{V \gamma/q_{i}q_{j}}
\end{equation}

where $\xi = \tau/\tau^\prime$ and

\begin{equation}
\Bigl.\frac{dL}{d\xi}\Bigr|_{V\gamma/q_{i}q_{j}} = \int_{\xi}^{1}
\frac{dx}{x}f_{V/q_{i}}(x) f_{\gamma/q_{j}}(\xi/x)
\end {equation}
In the above expression $f_{q/p}$ represents the quark structure functions of
the proton, which we have chosen to be the HMRS (Set B) structure
functions\cite{hmrs}. For the case $V = \gamma$ an analytic expression can be
obtained for the subprocess differential luminosity as

\begin{equation}
\Bigl.\frac{dL}{d\xi}\Bigr|_{\gamma\gamma/q_{i}q_{j}} =
\frac{e^2_{q_i}e^2_{q_j}}{64\pi^4}\left\{(4 - \frac{6}{\xi}+2\xi) -
(4+\frac{4}{\xi}+\xi)\log \xi\right\}\Bigl(\log \frac{t_{\rm max}}
{t_{\rm min}}\Bigr)^2
\end{equation}
No such simple analytic expression is available for the case $V = Z$ using
non-leading order $Z$ distribution functions.

The $\gamma \gamma$ differential luminosity for elastic $pp$ collisions is
given by,

\begin{equation}
\Bigl.\frac{dL^{\rm el}}{d\tau}\Bigr|_{\gamma\gamma/pp} = \int_{\tau/x_{\rm
 max}}^{x_{\rm max}} \frac{dx}{x}f^{\rm el}_{\gamma/p}(x) f^{\rm el}_
{\gamma/p}(\tau/x)
\end{equation}
where the upper limit of integration is given by kinematical considerations to
be $x_{\rm{max}} = (1 - 2m_p/\sqrt{s})$, $\sqrt s$ being the center of mass
energy of the elastically colliding protons.

The $V \gamma$ $(V = \gamma, Z)$ differential luminosity for semi-elastic $pp$
collisions is given by,

\begin{equation}
\Bigl.\frac{dL^{\rm{semi-el}}}{d\tau}\Bigr|_{V \gamma/pp} =
2\int_{\tau/x_{\rm {max}}}^{x_{\rm {max}}}\frac{dx}{x}f^{\rm el}_{\gamma/p}(x)
f^{\rm inel}_{V/p}(\tau/x)
\end{equation}
where

\begin{equation}
f^{\rm inel}_{V/p}(x) = \sum_{i}\int^1_{x_{\rm
min}}\frac{dx^\prime}{x^\prime}f_{q_i/p}(x^\prime)f_{V/q_i}(x/x^\prime)
\end{equation}

We present $\gamma \gamma$ luminosities in Fig. 2, $Z
\gamma$ luminosities derived with the non-leading Dawson effective $Z$
approximation in Fig. 3, and $Z \gamma$ luminosities derived with the Johnson
 {\sl et. al.} effective $Z$ distributions in Fig. 4. We consider these results
in turn below. The luminosity for the elastic case is independent of the
production process,   however for the other two cases it depends on the
production process through the factor $t_{\rm {max}}$.  In each Figure, the
luminosities for inelastic and semi-elastic cases are presented for the case
of charged lepton mass $m_{L} = 200$ $\rm GeV$.

In Fig. 2 we show the inelastic, elastic, and semi-elastic two photon
luminosities in $pp$ collisions for the
proposed CERN Large Hadron Collider (LHC) of center of mass energy of 14 TeV.
For comparison purposes we also show the two gluon luminosity.
As can be seen from Fig. 2 the $\gamma \gamma$ luminosities are comparable for
the three cases but are, however, 3-4 orders of magnitude smaller than the
$gg$ luminosity. This can be understood as being due to the fact that the
photon
luminosities are suppressed by a factor $(\alpha/\alpha_s)^2$ with respect to
the gluon luminosity, which is somewhat countered by the logarithmic
enhancement factor present in the photon luminosities.

In Figs. 3 a, b  we show the $Z \gamma$
differential luminosities for inelastic and semi-elastic $pp$ collisions,
respectively, using the non-leading Dawson form of the effective $Z$
approximation. The solid line corresponds to the luminosity for
longitudinally polarized $Z$ bosons, the dot-dashed line to the average over
the
two transverse polarizations of the $Z$. The luminosity corresponding to
transversely polarized $Z$ dominates that for longitudinal $Z$
polarizations. This may be understood as being due to helicity suppression.
For both the inelastic and semi-elastic cases, the      $Z_{T} \gamma$
differential luminosity is larger than the $Z_{L} \gamma$ differential
luminosity by a factor of 2-3 for $\tau \approx 10^{-3}$ and by   an order of
magnitude for $\tau \approx 10^{-1}$. Ultimately, however, we will see that the
$Z_L \gamma \rightarrow L^+L^-$ cross-section dominates so it is most important
to consider the luminosity for longitudinal $Z$ bosons here. Comparing the
inelastic and semi-elastic cases, the total differential  $Z \gamma$ luminosity
for the inelastic case is larger than that for the semi-elastic case by a
factor
of 2-3 for $\tau \approx 10^{-3}$, but for $\tau \geq 0.03$ they become equal.

In Figs. 4 a, b we show the results for the $Z \gamma$
differential luminosities, for inelastic and semi-elastic $pp$
collisions, respectively, using the Johnson form of the effective $Z$
approximation. Here, again, the solid line shows the luminosity for
longitudinally polarized $Z$ bosons. The dot-dashed and dotted lines show
separately the luminosities for positive and negative helicity transverse
Z polarizations, respectively. The dashed line corresponds to the sum of the
luminosities for the two transverse Z polarizations, which we will refer to as
$Z_{T}$ as in the Dawson case. One observes essentially the same numerical
relationships between the different luminosities shown here, as discussed above
for the Dawson form of the effective $Z$ approximation. The $Z_{L} \gamma$
luminosities agree almost exactly for the two different forms of the effective
$Z$ approximation, except at very small $\tau$.  The Johnson distribution
functions yield $Z_{T} \gamma$ luminosities slightly larger than those obtained
with the Dawson distribution functions. We will find that the $Z \gamma$
process cross section is only large enough to be interesting for heavy lepton
masses above about 600 $\rm GeV$; in the range of $\tau$ corresponding to
production of such heavy leptons the Dawson and Johnson luminosities for
$Z_{T}$ agree to 20 \% or better. In both the longitudinal and transverse
cases, one does expect some disagreement between the Dawson and Johnson
formulations. While they are both non-leading distribution functions, they are
derived under quite different kinematical assumptions; the Dawson derivation
includes approximations, such as the small-angle one, which the Johnson
derivation does not. In addition, for the transverse $Z$ polarizations, the
Johnson formulation separately extracts the positive and negative $Z$ helicity
distribution functions whereas the Dawson formulation extracts a single
transverse $Z$ distribution function from the cross sections averaged over the
two transverse polarizations. So, particularly in the transverse case, the
agreement should not be perfect. We have shown in Figs. 3 and 4 a detailed
comparison of the luminosities   using the two forms of the effective $Z$
approximation. We will present the  production cross-sections for $Z\gamma$
fusion using only the non-leading order Dawson form of the effective $Z$
approximation in the next Section. This choice is really rather arbitrary. For
the longitudinal $Z$ case, which will be of interest, agreement between the
two formulations is excellent.

Comparing the two different initiating states, we find that the $Z \gamma$
luminosities are smaller than the $\gamma \gamma$ luminosities for small values
of $\tau$, but they do not     differ by more than an order of magnitude.
For $\tau \geq 10^{-2}$ the $Z \gamma$ and $\gamma \gamma$ luminosities
are of the same
order of magnitude.

\section{The $L^+L^-$ Production Cross Sections}

The matrix elements and
cross-sections for the relevant subprocesses, $V \gamma \rightarrow L^+L^-$,
are
given in the Appendix. The total
cross-sections are obtained by convoluting
the subprocess cross sections
with the appropriate photon (and/or $Z$) and quark structure functions and
integrating
over the two body phase space using Monte Carlo techniques.

The cross-section for the $V \gamma$ production of a pair of heavy charged
leptons in inelastic $pp$ collisions (with center of mass energy $\sqrt s$)
is obtained by
convoluting the  cross-section of the $V \gamma \rightarrow
L^+L^-$ subprocess, $\sigma_{V \gamma}$, with the
probabilities of finding the gauge bosons in the protons as follows.

\begin{equation}
\sigma^{\rm inel} (s) = \int_{x_{\rm min}}^{1} dx_1
\int_{x_{\rm min}/x_1}^{1} dx_2 \, f^{\rm inel}_{V / p} (x_1) \,
f^{\rm inel}_{\gamma / p} (x_2) \, \sigma_{V \gamma}(x_1 x_2 s)
\end{equation}
Here $V$ represents either a photon or a longitudinal or transverse degree of
freedom of the $Z$ boson.
The inelastic component of the $V$ spectrum from protons,
$f^{\rm inel}_{V / p} (x)$, is given by

\begin{equation}
f^{\rm inel}_{V / p} (x) = \int_{x_{\rm min}}^{1} dx_1
\int_{x_{\rm min}/x_1}^{1} dx_2 \, \sum_{q}
f_{V / q} (x_1) \, f_{q / p} (x_2, Q^2) \, \delta (x - x_1 x_2)
\end{equation}

\noindent
In the above equation, the $f_{q/p} (x_2, Q^2)$ represent the quark
structure functions for the proton\cite{hmrs}, evaluated at the scale
$Q^2 = \hat{s}/4$, $\hat{s}$
being the parton center of mass energy squared. We sum over the contributions
due to the $u, d, c$, and $s$ quarks and antiquarks from the protons.
The lower limit of integration in the above equations ensures that
the $V \gamma$ center of mass energy is sufficient for $L^+L^-$ production.

The cross-section for two photon $L^+L^-$ production via elastic collisions
of protons
is obtained by folding the $\gamma \gamma$ subprocess cross-section,
$\sigma_{\gamma    \gamma}$, with
the elastic component of the photon spectrum from the protons, $f^{\rm el}_
{\gamma / p}$.

\begin{equation}
\sigma^{\rm el} (s) = \int_{x_{\rm min}}^{x_{\rm max}} dx_1 \int_{x_{\rm
min}/x_1}^{x_{\rm max}} dx_2 \, f^{\rm el}_{\gamma / p}
(x_1) \,
f^{\rm el}_{\gamma / p} (x_2) \, \sigma_{\gamma \gamma}(x_1 x_2 s)
\end{equation}

\noindent The upper limits of integration in the above expression are given by
$x_{\rm max} = 1 - 2 m_p/{\sqrt s}$, where $\sqrt s$ is
the center of mass energy of the elastically colliding
protons. The lower limit of integration $x_{\rm min}$ has the same significance
as in the case of inelastic collisions.
We have checked this calculation by
using an alternative form of the photon spectrum from elastically colliding
protons\cite{dhkz}. The cross sections obtained with this alternative form
essentially agree with the results presented here using the Kniehl form of WWA.

The cross sections for $L^+L^-$ production via semi-elastic collisions of
protons, induced by $\gamma \gamma$ and $Z \gamma$ are obtained in a similar
fashion by appropriately combining the two cases detailed above.

For comparison with the mechanisms considered here, we also
present results for the two previously well known production mechanisms for
$L^+L^-$, Drell-Yan and gluon fusion.
The Drell-Yan quark anti-quark annihilation proceeds
via s-channel $\gamma$ and $Z$ exchange. The gluon fusion proceeds via a quark
triangle diagram followed by $Z$ boson or Higgs boson exchange. Expressions for
these
cross sections can be found in the literature\cite{gf,barn}. The rates for
the other possible processes of weak gauge boson fusion and
annihilation are negligible compared to the Drell-Yan
and gluon fusion rates except for the case of very massive heavy leptons,
\cite{gf} and will not be discussed further.

In our calculations we use $M_W = 80.22$ $\rm GeV$, $M_Z = 91.19 \,$
$\rm GeV$. In calculating the two photon production cross-sections we have
taken  into account the variation of the fine structure constant $\alpha$ with
the energy scale, {\sl e.g.}, we have chosen $\alpha = 1/137$ in the
expressions
for Weizs\"{a}cker-Williams photon spectrum, and $\alpha = 1/128$ in the
expressions for the subprocess cross-sections.
The parameter $\alpha_s$, used in the gluon fusion
calculation, is evaluated at the two-loop
level using its representation in the modified minimal-subtraction $(\overline
{\rm MS})$ scheme\cite{alfs}, and using the value $\Lambda_{\overline {\rm MS}}
^{(4)} = 0.19$ $\rm GeV$, consistent with the HMRS (Set B) parametrization.
For the gluon fusion mechanism we assume three
generations of quarks, with the top quark mass set at 175 $\rm GeV$. Results
for    the inclusion of a fourth generation of heavy quarks can be found in
Refs. 2,   19. The Higgs boson mass for our calculation is chosen to be 200
$\rm GeV$. There is very little sensitivity to this parameter.

The total cross sections for $L^+L^-$ production via two photon fusion
as a function of the charged lepton
mass, are shown in Fig. 5 for the LHC center of mass energy of 14 TeV.
The dotted and upper solid curves represent respectively the cross sections
due to gluon fusion and Drell-Yan mechanisms.
The lower solid curve represents the total cross section for the
 inelastic process $p p \rightarrow \gamma \gamma X \rightarrow
L^+L^- X$. The dashed and dot-dashed curves show the corresponding cross-
sections for the elastic and semi-elastic cases respectively. The cross-
sections for the three cases are comparable to each other over the mass range
of the charged lepton. However they are 2-3 orders of magnitude smaller than
the Drell-Yan
and gluon fusion cross sections.

Hence in general, production rates via two photon fusion at $pp$ colliders are
not competitive with production via quark or gluon fusion. Only the
interesting event topology of particle production through photon fusion
for elastic and semi-elastic $pp$ collisions might offer a chance of
distinguishing signals which are otherwise very difficult in a hadronically
noisier  environment. Both the elastic and semi-elastic processes should yield
clean events with the intact proton(s) continuing in the forward direction.
However, the low event rates probably eliminate this as a viable process for
$L^+L^-$ discovery. Two recent publications\cite{do} address similar topics for
two photon processes as discussed here, and reach essentially the same
conclusions. Drees  {\sl et al.} point out the additional problem of multiple
events at high luminosity; thus, the high luminosity necessary to achieve an
interesting rate will destroy the cleanliness of the signal.

We now turn to the $L^+L^-$ production via $Z \gamma$
fusion. In the Appendix, we present the matrix element squared for the
$Z \gamma$ subprocess, without summing over the $Z$ boson polarization. This
allows us to obtain separately the contributions from longitudinally and
transversely polarized $Z$ bosons. We also give an expression for the
subprocess cross-section summed over $Z$ boson polarizations in the Appendix.

For subprocess energies much greater than the $Z$ boson
mass the interactions between the longitudinally polarized $Z$ boson and
fermions can be derived in terms of an effective theory\cite{effth} in which
$z_L$ is the would be Goldstone boson of the broken $SU(2)_L \times U(1)_Y$
model. Calculations are greatly simplified in the effective theory since
the vector bosons are replaced by scalar bosons. Hence as an additional check
on our exact calculation we have calculated the
cross-section for the subprocess $Z_L \gamma \rightarrow L^+L^-$ in the
Goldstone boson approximation. The subprocess cross-section for longitudinal
$Z$ bosons dominates
since the coupling of fermions to longitudinally polarized $Z$ bosons is
proportional to the large fermion mass in the effective Lagrangian, as given
by Dawson\cite{gbf}

\begin{equation}
{\cal L} \sim \frac{ig}{2M_Z \cos\theta_W} m_L\bar l \gamma_5 l z_L
\end{equation}
Here $l$ indicates the heavy charged lepton wave function and $z_L$ that of the
Goldstone boson. The subprocess cross-section is given in the Appendix. Indeed
we find good agreement between this approximate calculation and our
results obtained from the exact calculation.

In Figs. 6 a, b we show respectively the $L^+L^-$ production cross-sections via
$Z \gamma$ fusion at the LHC for inelastic and semi-elastic $pp$ collisions. We
have used the non-leading Dawson form of the $Z$ distribution function. For
reference, the cross
sections due to gluon fusion and Drell-Yan mechanisms are again shown as the
dotted and upper solid curves, respectively. The dashed and dot-dashed
curves represent, respectively, the separate contributions of longitudinally
and
transversely polarized  $Z$ bosons. The solid curve is the total cross section.
The $Z_{T}\gamma$ cross-section is smaller than that for $Z_{L} \gamma$ and
also
decreases more rapidly with increasing $m_L$. For small
values of $m_L$ the cross-sections for semi-elastic $pp$ collisions are smaller
than those for inelastic collisions, but they become approximately equal
for $m_L \geq 500$ $\rm GeV$.

Our results for $Z\gamma$ production of heavy charged lepton are in
disagreement with those of Eboli {\sl et al.}\cite{eboli} Those authors
consider inelastic $Z \gamma$ production in $pp$ collisions using the leading
logarithmic result for the effective $Z$ approximation, and using the
Goldstone boson approximation for the subprocess cross-section. However, as has
been pointed out in Ref. 6, they have chosen the $Z_L$ coupling to quarks in
the effective Lagrangian to
be proportional to the vector coupling of $Z$ bosons to quarks,
whereas it should be proportional to the axial vector coupling.
They find
the transversely polarized $Z$ boson cross-sections dominating the
longitudinally polarized $Z$ boson cross-section. While we agree that the
$Z_T \gamma$ luminosity is larger than that for $Z_L \gamma$, we find that the
$Z_L \gamma \rightarrow L^+L^-$ cross-section dominates. As noted
above we have good agreement between our exact calculation and the
Goldstone boson approximation calculation for the $Z_L \gamma$ subprocess.

One can see that the $Z \gamma$ inelastic cross-sections are comparable to the
$\gamma \gamma$ inelastic cross-sections for small values of $m_L$. However
the $\gamma \gamma$ fusion cross-section falls off much more rapidly with
increasing $m_L$ compared to the $Z \gamma$ fusion cross-section, and for
larger values of $m_L$ the $Z \gamma$ cross-section is an order of magnitude
larger than the $\gamma \gamma$ cross-section. One may intuitively understand
the relative magnitudes of the $\gamma \gamma$ and $Z \gamma$ cross-sections
as follows. The ratio of the subprocess cross-sections for $Z \gamma
\rightarrow L^+L^-$ and $\gamma \gamma \rightarrow L^+L^-$ for the same
subprocess center of mass energy (i.e. same $\tau$) is almost independent
of $\tau$ but depends only on $m_L$. The $Z \gamma$ subprocess cross-section
is approximately an order of magnitude larger than that for $\gamma \gamma$
for $m_L \approx 200$ $\rm GeV$ and is about two orders of magnitude larger for
$m_L \approx 700$ $\rm GeV$. Since the $Z \gamma$ differential luminosities in
$pp$ collisions are about an order of magnitude smaller than the $\gamma
\gamma$ differential luminosities for small $\tau$, and approximately equal
for $\tau \geq 10^{-2}$, the relative magnitudes of the total cross-sections
obtained in $pp$ collisions follow.

The result is that each of the inelastic and semi-elastic $Z \gamma$
production cross sections for $L^+L^-$ is within about an order of magnitude
of the Drell-Yan and gluon fusion cross sections for large values of $m_{L}$,
beyond $m_{L}$ about 500 $\rm GeV$ in the inelastic case and somewhat higher in
the semi-elastic case. For these large lepton masses, the cross sections for
the $Z \gamma$ processes are down at the tenths of a fb level, so high
luminosity is again important. A one year integrated luminosity of 100
${\rm fb}^{-1}$ provides about 20 pairs of 600 $\rm GeV$ leptons for each of
the
inelastic and semi-elastic mechanisms.

We now discuss qualititatively the possible signatures for $L^+L^-$ pair
production at $pp$ colliders via $Z\gamma$ fusion and their potential
backgrounds. We have discussed the production of $L^+L^-$ so far in a
relatively model-independent fashion. The production cross-sections for
heavy charged leptons that are predicted within a large variety of models,
are either given directly by the cross-section results presented here or can
be obtained by a simple rescaling of these results. The decays for heavy
charged leptons and their possible signatures depend however on other extra
particles predicted within the model. Thus a model independent discussion of
the observability of $L^+L^-$ pairs is not possible. Previous work on
detailed analysis of $L^+L^-$ signal and backgrounds by Barger {\it et. al.}
\cite{gen} and I. Hinchliffe\cite{hinch} assumed a sequential (4th generation)
charged heavy lepton and its associated neutrino $\nu_L$. For simplicity we
consider the same model here in our discussion of the possible signatures.
Depending on the decay mode of charged heavy leptons there can be three
possible signatures which we discuss in turn.

If both charged heavy leptons decay leptonically, {\it e.g.} $L^+ \rightarrow
\bar\nu_L \bar l \nu_l$, $L^- \rightarrow \nu_L l^\prime \bar{\nu_{l^\prime}}$
$(l, l^\prime = e, \mu)$, the signature would be an $e^+e^-$, or $\mu^+\mu^-$,
or $e^\pm \mu^\mp$ pair and missing transverse momentum. The SM backgrounds
in this case are: (a) $pp \rightarrow Z^*\gamma^* \rightarrow \tau \bar\tau$
(b) $pp \rightarrow W^+W^-$ (c) $pp \rightarrow ZZ$ (d) $pp \rightarrow t \bar
 t \rightarrow bW^+\bar bW^-$ where the gauge bosons and $\tau$ leptons
themselves decay
leptonically. There are also additional backgrounds to be considered, the
two-photon fusion processes at $pp$ colliders, $\gamma \gamma \rightarrow
W^+W^-
$ and $\gamma \gamma \rightarrow l \bar l$.
If one heavy lepton decays leptonically and the other hadronically, {\it e.g.}
$L^+ \rightarrow \bar \nu_L q \bar q^\prime$, $L^- \rightarrow \nu_L l \bar
\nu_l$, the final state consists of an electron or muon, missing $p_T$ and
jets. The SM backgrounds for this signature are : (a) $pp \rightarrow W^- +$
jets (b) $pp \rightarrow W^+W^-$ (c) $pp \rightarrow t \bar t \rightarrow
bW^+\bar b W^-$, and the process $\gamma\gamma \rightarrow W^+W^-$, where for
the cases of $W$ pair production, one
$W$ boson of the pair decays leptonically and the other hadronically.
If both charged heavy leptons decay hadronically the signature is 4 jets and
missing $p_T$. The SM background in this case is $pp \rightarrow Z +$ 4 jets,
with $Z \rightarrow \nu \bar \nu$.

Detailed Monte-Carlo simulations of the signals and backgrounds
for these decay modes of $L$, but corresponding to a
different production mode, in inelastic $pp$ collisions, have been reported
previously\cite{gen,hinch}. For the semi-elastic $Z \gamma$ fusion case,
the signals corresponding to the various decay modes of $L$ considered above
would include the additional tag of a single proton. In principle
the unique topology of the semi-elastic $Z \gamma$ events could be exploited to
observe the signal, for example by using forward spectrometers. The results
presented here are intended as an exploratory survey of the novel signatures
that occur in photon and $Z$ induced fusion processes at hadron colliders.
Detailed confirmation of the observability of these signatures would require
Monte-Carlo simulations of the signals and backgrounds, including considering
the detector acceptance, and effects such as overlapping events at high
luminosities.

\section{Conclusions}

We have evaluated the importance of hadron colliders as a source of
$\gamma \gamma$ and $Z \gamma$ collisions by considering the production of
charged heavy leptons in such colliders. While we have used the pair production
of charged heavy leptons as a benchmark process, some of our results are of
broader interest. In particular, we have presented $\gamma \gamma$ and $Z
\gamma$ luminosities for inelastic, elastic, and semi-elastic $pp$ scattering.
The luminosities for longitudinally and transversely polarized $Z$ bosons were
given separately. These results can be used for the consideration of the
production of other final states via the $\gamma \gamma$ and $Z \gamma$
initiating subprocesses.

Our conclusions with respect to the two photon induced heavy lepton
production are rather pessimistic, consistent with Drees  {\sl et al.} \cite
{do}.  The total event rates for
$\gamma \gamma$ induced processes are  not competitive with other dominant
production mechanisms, being down by a couple orders of magnitude. The
novel mechanism of obtaining photons without one or both of the protons
breaking up could, in principle, prove useful in confirming signals which are
difficult to otherwise distinguish. However, the $\gamma \gamma$ cross sections
are probably too suppressed to exploit the topology of the elastic events.

On the other hand, we find that $Z \gamma$ production of heavy charged leptons
overtakes  $\gamma \gamma$ production for lepton masses above 100 $\rm GeV$. It
rises to within about an order of magnitude of the dominating Drell-Yan and
gluon
fusion processes for $m_L$ of about 600 $\rm GeV$. This is the same sort of
behaviour as seen in Higgs boson production via the gluon and vector boson
fusion
mechanisms, where the vector boson fusion cross-section ultimately overtakes
the gluon fusion cross-section for large Higgs mass\cite{cahn}. Hence the $Z
\gamma$ initiated processes should not be neglected. The inelastic and
semi-elastic processes together make up about 10 \% of the total event rate for
$m_{L}$ of 600 $\rm GeV$. The cleaner semi-elastic production could in
principle
stand out as a distinguishable signal, however direct confirmation of that
requires detailed detector simulation studies.

%Finally the consideration of
%these processes serves as an important testing ground for the
%effective photon and $Z$ approximations that were used to evaluate
%the production rates.

\acknowledgements
This work was funded in part by the Natural Sciences and Engineering Research
Council of Canada. The authors acknowledge useful conversations with Rohini
Godbole, Manuel Drees and Dieter Zeppenfeld.

\appendix
\section*{}
We present here expressions for the squared matrix elements and subprocess
cross-sections for $L^+L^-$
pair production through $\gamma\gamma$ and $Z \gamma$ fusion.

Assuming that heavy charged leptons couple to the photon in
the usual way, the summed and averaged matrix element squared for the
2 photon fusion subprocess is given by

\begin{eqnarray}
\sum \overline{|{\cal M_{\gamma\gamma}}|^2} & = & 2 e^4 \biggl[\frac{(u-m_L^2)}
{(t-m_L^2)} + \frac{(t-m_L^2)}{(u-m_L^2)}
- 4 m_L^2 \Bigl(\frac{1}{(t-m_L^2)}+\frac{1}{(u-m_L^2)}\Bigr) \nonumber \\
& &- 4 m_L^4 \Bigl(\frac{1}{(t-m_L^2)} +\frac{1}{(u-m_L^2)}\Bigr)^2\biggr]
\end{eqnarray}

\noindent where $t$ and $u$ refer to the exchanged momenta squared
corresponding to
the direct and crossed diagrams (Figs. 1 a, b) for the two photon
subprocess, and the photon coupling is represented by $e^2 = 4\pi\alpha$,
with the value of the fine structure constant $\alpha = 1/128$ which is
appropriate for high energy production processes. The total subprocess
cross-section for $\gamma\gamma \rightarrow L^+L^-$ is given by

\begin{equation}
\sigma(\gamma\gamma \rightarrow L^+L^-) = \frac{e^4}{4\pi
s^2}\left[\left(\frac{8m_L^4}{s} - 4m_L^2
-s\right)\log\left(\frac{1-\beta}{1+\beta}\right) - \beta(s+4m_L^2)\right]
\end{equation}
where $\beta = (1 - 4m_L^2/s)^{\frac{1}{2}}$, $s$ being the $\gamma\gamma$
center of mass energy.

We present the matrix element squared for the $Z \gamma$ subprocess
after summing and averaging over the photon polarization but not the
$Z$ polarization.

\begin{equation}
\sum_{\gamma} \overline{|{\cal {M}}_{Z\gamma}|^2} = \frac{e^4}
{x_W(1-x_W)}
\biggl[\frac{T_1}{(t-m^2_L)^2}+\frac{T_2}{(u-m^2_L)^2}+\frac{T_3}
{(t-m^2_L)(u-m^2_L)}\biggr]
\end{equation}
where

\begin{eqnarray}
T_1 & = & 2\Bigl[(g^2_V+g^2_A)\left\{(t-u)m^2_L + ut - 4Q^{\prime 2} (t-3
m^2_L)
- 4 Q Q^\prime (t-m^2_L) - 2 M^2_Zm^2_L + 3 m^4_L\right\} \nonumber\\
& & - \; 2(g^2_V-g^2_A)\left\{m_L^2(t + m_L^2)\right\}\Bigr] \\
T_2 & = & T_1 (Q \leftrightarrow  Q^\prime, t \leftrightarrow u) \\
T_3 & = & 4\Bigl[(g^2_V+g^2_A)\left\{(t+u)(m^2_L-M^2_Z)
+ 2 Q^2(t+m^2_L-M^2_Z) + 2Q^{\prime 2}(u+m^2_L-M^2_Z)\right. \nonumber\\
& &  \left. + \;2QQ^\prime(t+u-2m^2_L) - 2m^2_LM^2_Z +
2m^4_L + M^4_Z\right\} \nonumber\\
& & - \;(g^2_V-g^2_A)\left\{m^2_L(4Q^2 + 4Q^{\prime 2} + 8QQ^\prime
+ t + u - 2M^2_Z) + 2m^4_L\right\}\Bigr]
\end{eqnarray}
In the above expressions $g_V = (-\frac{1}{4} + x_W)$ and $g_A = \frac{1}{4}$
are respectively the vector and axial vector
couplings of the heavy lepton to $Z$ boson, $x_W = \sin^2 \theta_W$, $\theta_W$
being the Weinberg angle, and

\begin{equation}
Q = q \cdot \epsilon _{Z}, \; \; \; Q^\prime = q^\prime \cdot \epsilon_{Z}
\end{equation}
where $\epsilon_{Z}$ is the $Z$ polarization vector and $q, q^\prime$
are, respectively the $L^-$ and $L^+$ momentum 4-vectors.

Summing and averaging over $Z$ polarizations, the total subprocess
cross-section
for $Z\gamma  \rightarrow L^+L^-$ is given by
\begin{eqnarray}
\sigma(Z\gamma \rightarrow L^+L^-) & = & \frac{e^4}{6x_W(1-x_W)}\frac{8}{16\pi
(s-M_Z^2)^2}\left[(g_V^2-g_A^2)\left\{-\frac{12s\beta m_L^2}{s-M_Z^2} +
\frac{m_L^2}{s-M_Z^2}\left(2M_Z^2  \right.\right.\right. \nonumber\\
 & - & \left.\left. 16s + 24m_L^2 + 2\frac{s^2}{M_Z^2}\right)
{\rm log}\left(\frac{1-\beta}{1+\beta}\right)\right\} + (g_V^2+g_A^2)\left\{
-2\beta(s-M_Z^2) \right. \nonumber\\
 & + & 4s\beta\frac{m_L^2-M_Z^2}{s-M_Z^2} + {\rm log}
\left(\frac{1-\beta}{1+\beta}\right)\left(-2(s-m_L^2-M_Z^2)\right.\nonumber\\
 & + & \left.\left.\left. \frac{m_L^2}{s-M_Z^2}\left(6s + 8M_Z^2 - 8m_L^2 -
 \frac{4sM_Z^2}{m_L^2} - 2\frac{s^2}{M_Z^2}\right)\right)\right\}\right]
\end{eqnarray}
where $\beta = (1 - 4m_L^2/s)^{\frac{1}{2}}$, $s$ being the $Z\gamma$
center of mass energy.

The expression for the total cross-section for the subprocess $Z_L \gamma
\rightarrow L^+L^-$, in the Goldstone boson approximation, is given by:

\begin{eqnarray}
\sigma(Z_L\gamma \rightarrow L^+L^-) & = & \frac{e^4}{3x_W(1-x_W)}
\frac{m_L^2}{16\pi M_Z^2(s-M_Z^2)^2}\left\{M_Z^2\left(\frac{-2s\beta}{s-M_Z^2}
\right)\right. \nonumber\\
& - &  \left. 2{\rm log}\left(\frac{1-\beta}{1+\beta}\right)\left[\frac{1}{2}
(s-M_Z^2) + M_Z^2\left(\frac{s-2m_L^2}{s-M_Z^2}\right)\right]\right\}
\end{eqnarray}
%%%%%%%%%%%%%%%%%%%%%%%%%%%%%%%%%%%%%%%%%%%%

%%%%%%%%%%%%%%%%%%%%%%%%FIGURE CAPTIONS%%%%%%%%%%%%%%%%%%%%%%%%%%%
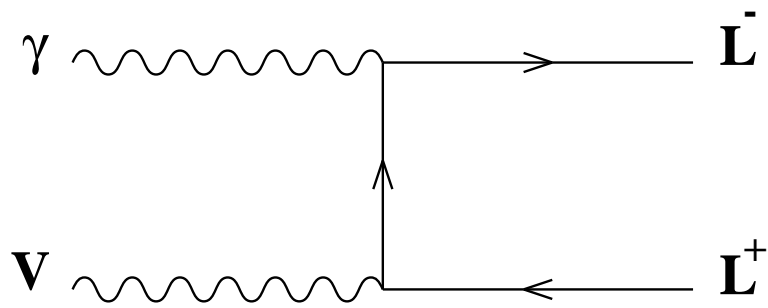
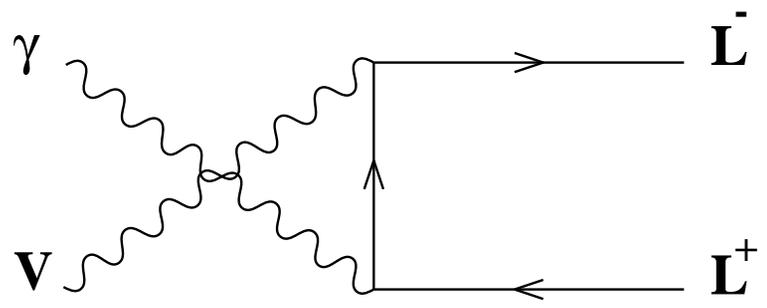
\begin{figure}
\caption[]{The Feynman diagrams which contribute to the subprocess
$V\gamma \rightarrow L^+L^-$ ($V = \gamma, Z$ ).
\label{fig1}}
\end{figure}

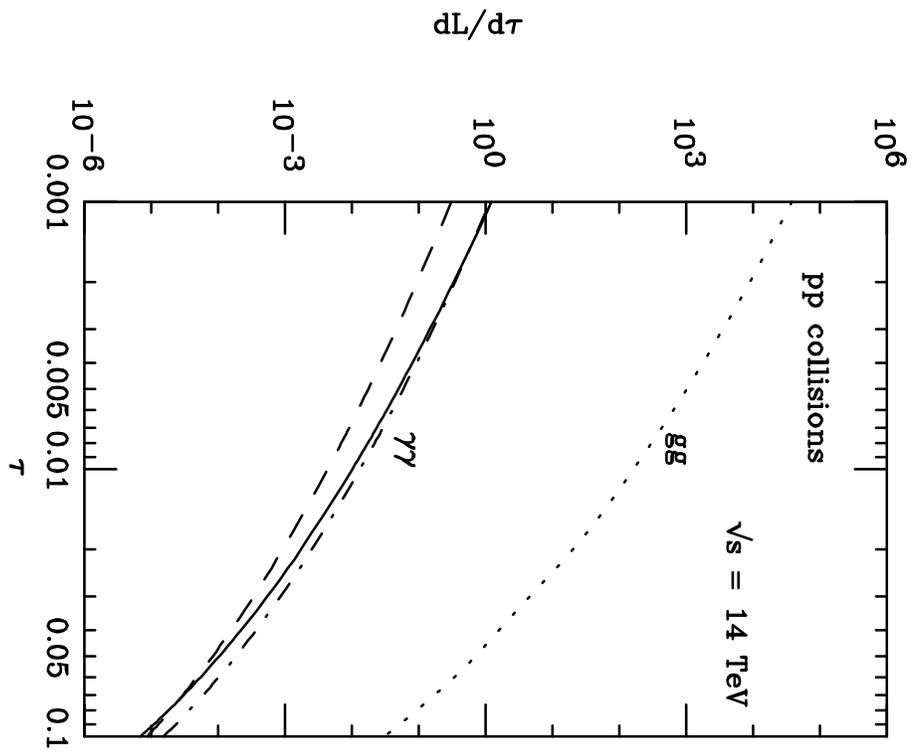
\begin{figure}
\caption[]{The differential $gg$ and $\gamma \gamma$ luminosities
in $pp$ collisions at the LHC ($\sqrt s = 14$ TeV).
The solid, dashed and dot-dashed curves represent respectively the
$\gamma \gamma$ luminosities for inelastic, elastic and semi-elastic
$pp$ collisions. The inelastic and semi-elastic luminosities are shown for the
case when the charged leptons produced via photon fusion has mass $m_L=$ 200
$\rm GeV$. The dotted curve shows the gluon luminosity.
\label{fig2}}
\end{figure}

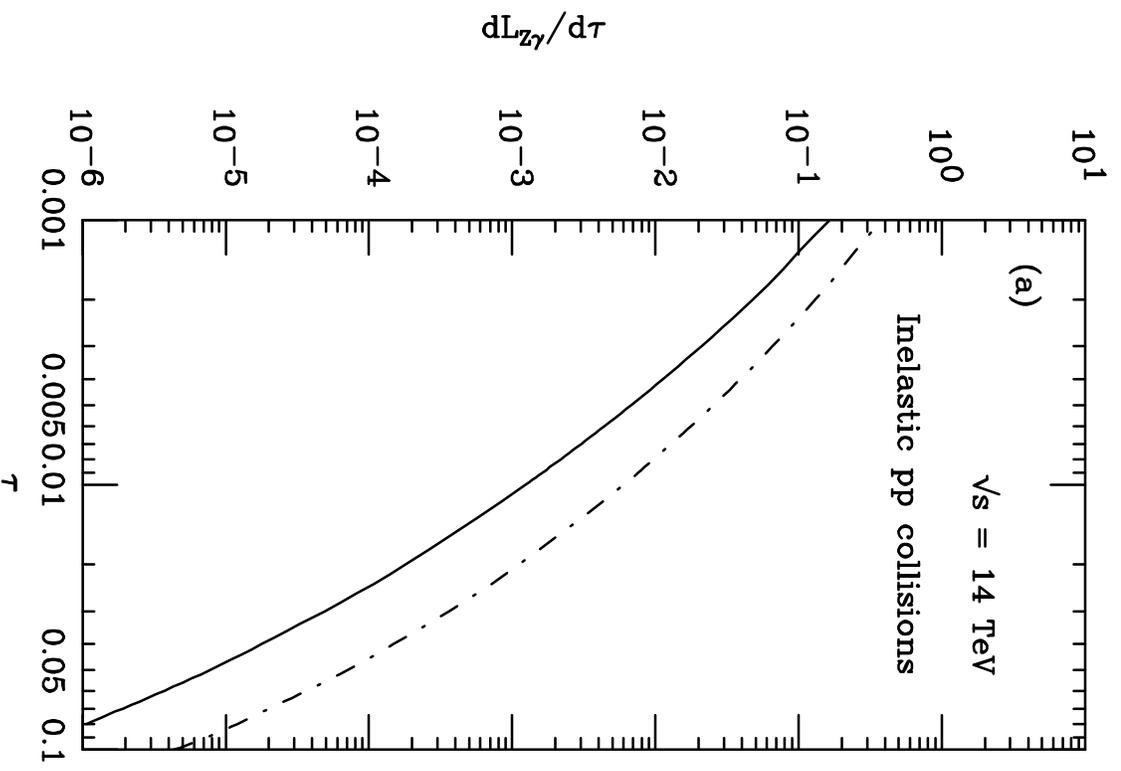
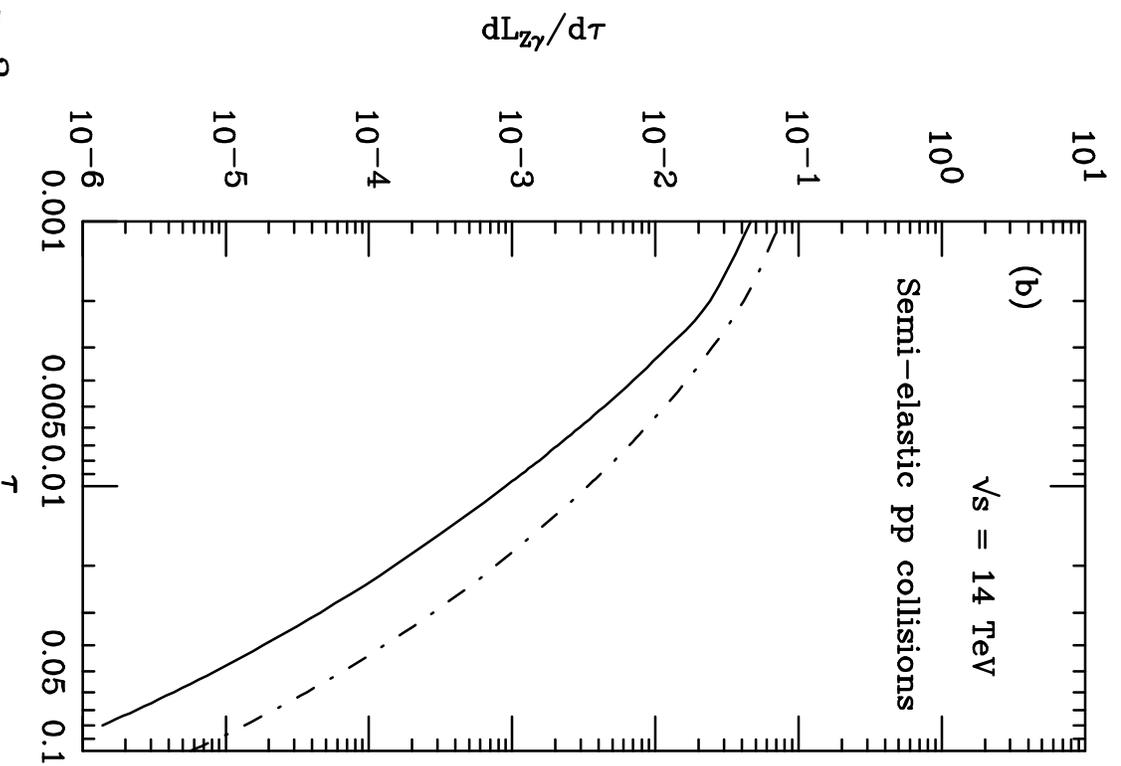
\begin{figure}
\caption[]{The differential $Z \gamma$ luminosities
in $pp$ collisions at the LHC ($\sqrt s = 14$ TeV) corresponding to the
non-leading Dawson  form of the effective $Z$ approximation. (a) The solid and
dot-dashed     curves represent, respectively, the $Z \gamma$ luminosities due
to the longitudinal and transverse $Z$ boson polarizations for inelastic
$pp$ collisions. (b) The curves have the same meaning as in (a)
except that they are for semi-elastic $pp$ collisions. The luminosities shown
in both figures are for the case when the charged leptons produced via $Z
\gamma$ fusion have mass $m_L=$ 200 $\rm GeV$.
\label{fig3}}
\end{figure}

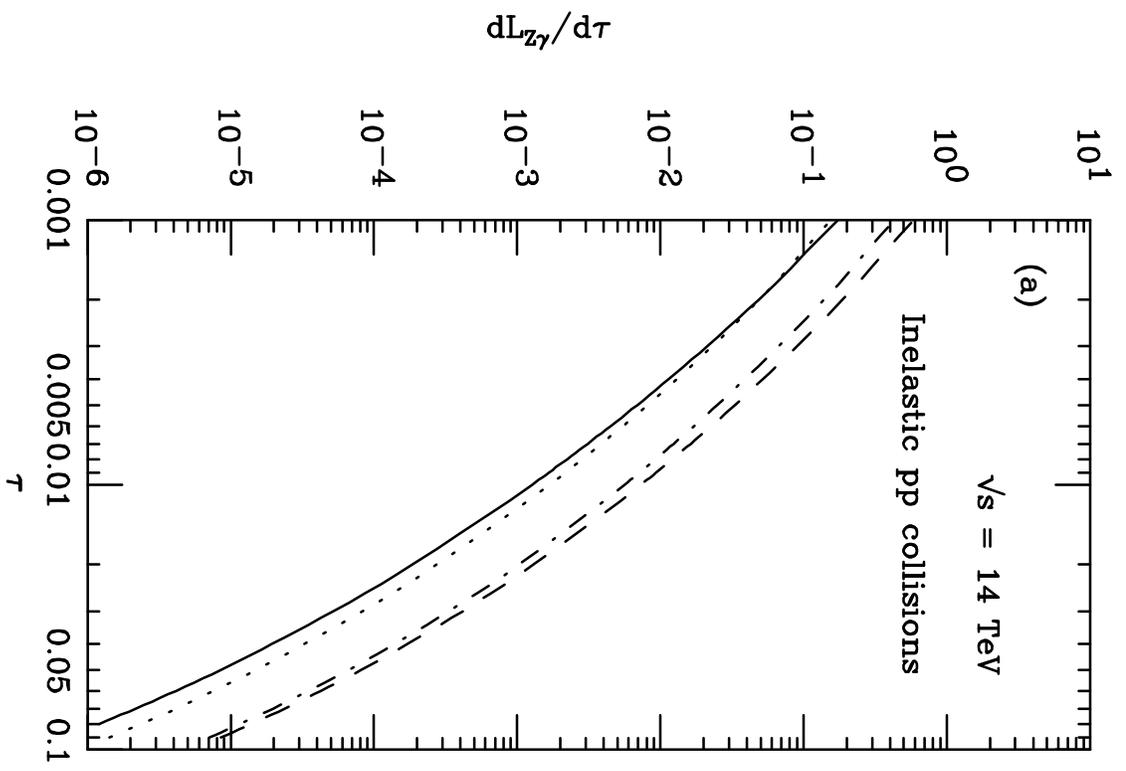
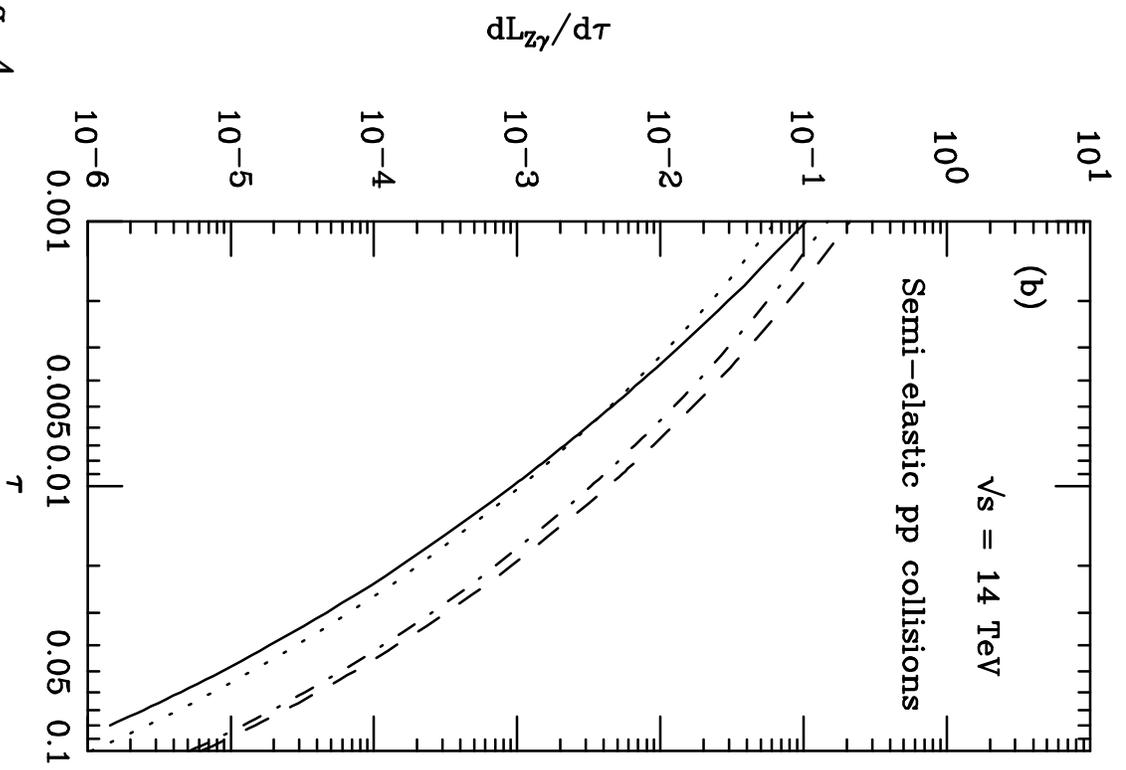
\begin{figure}
\caption[]{The differential $Z \gamma$ luminosities
in $pp$ collisions at the LHC ($\sqrt s = 14$ TeV) corresponding to the Johnson
form of the effective $Z$ approximation. (a) The solid, dot-dashed and
dotted curves represent respectively the $Z \gamma$ luminosities due to
the longitudinal, positive helicity, and negative helicity transverse
$Z$ boson polarizations for inelastic $pp$ collisions. The dashed
curve represents the luminosity for the sum of the two transverse $Z$
polarizations.  (b) The curves have the the same meaning as in (a) except
that they are for semi-elastic $pp$ collisions. The luminosities shown  in both
figures are for the case when the charged leptons produced via $Z
\gamma$ fusion have mass $m_L=$ 200 $\rm GeV$. \label{fig4}}
\end{figure}

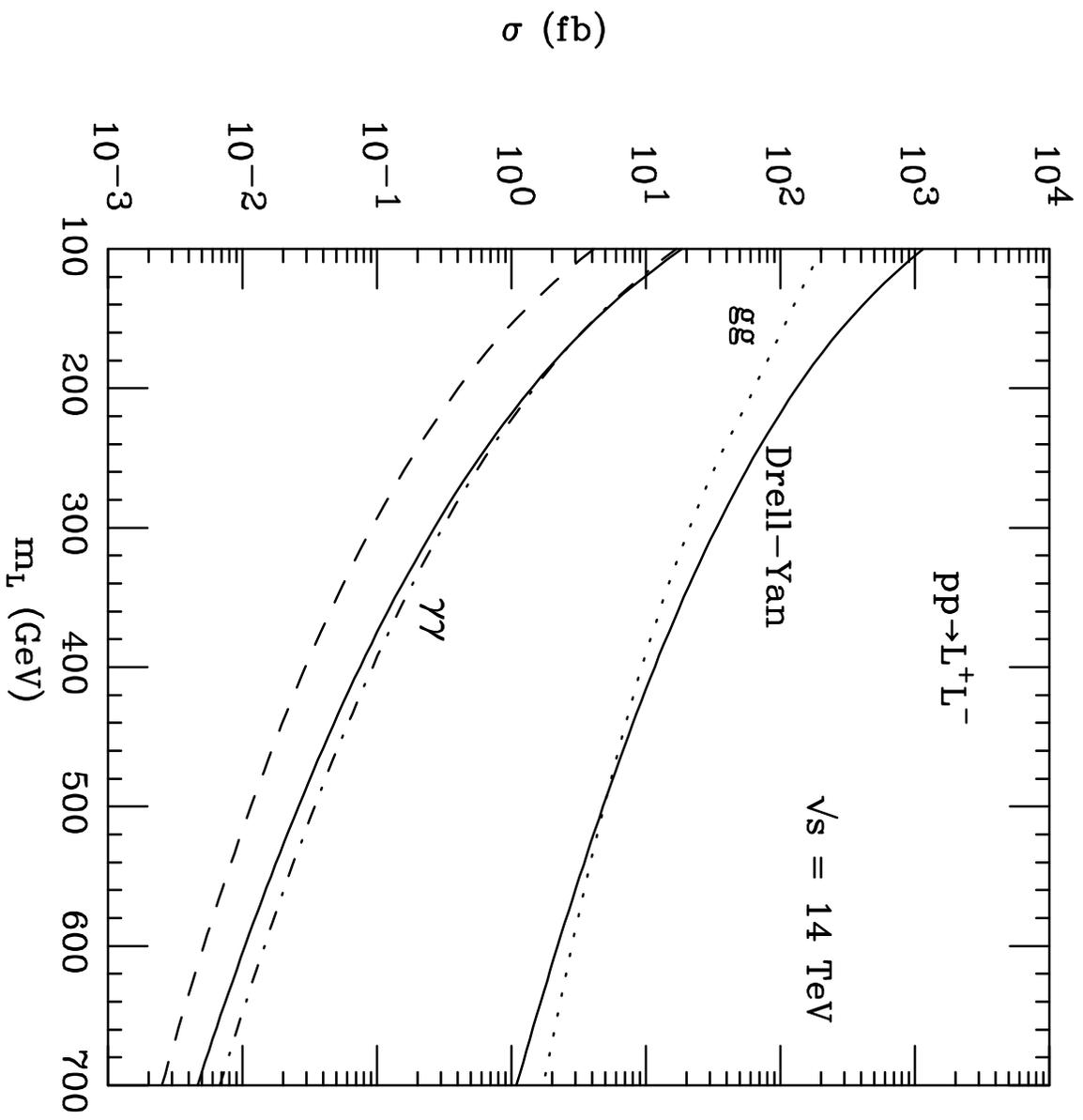
\begin{figure}
\caption[]{The total production cross section (in femtobarns) for a heavy
charged lepton pair in $pp$ collisions at the LHC ($\sqrt s = 14$ TeV).
The upper solid curve and the dotted curve show,
respectively, the Drell-Yan and gluon fusion cross sections. The lower solid
curve, dashed and dot-dashed curves represent respectively the photon
fusion cross-sections for inelastic, elastic and semi-elastic $pp$ collisions.
\label{fig5}}
\end{figure}

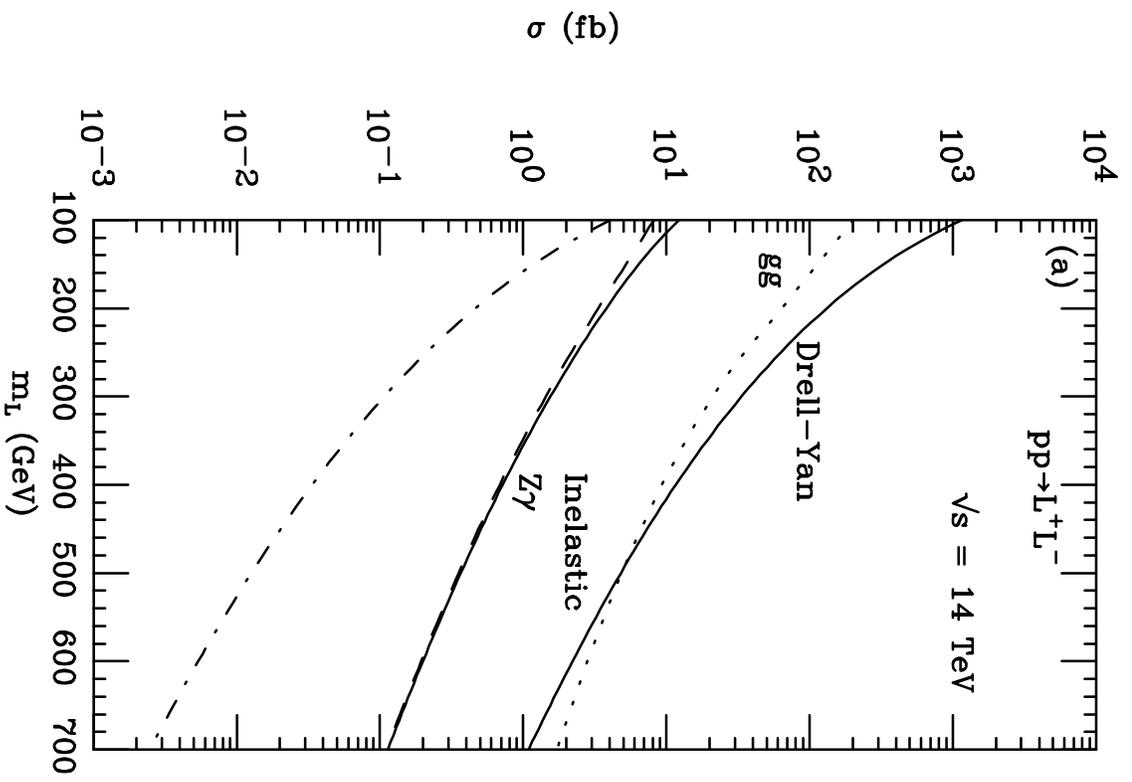
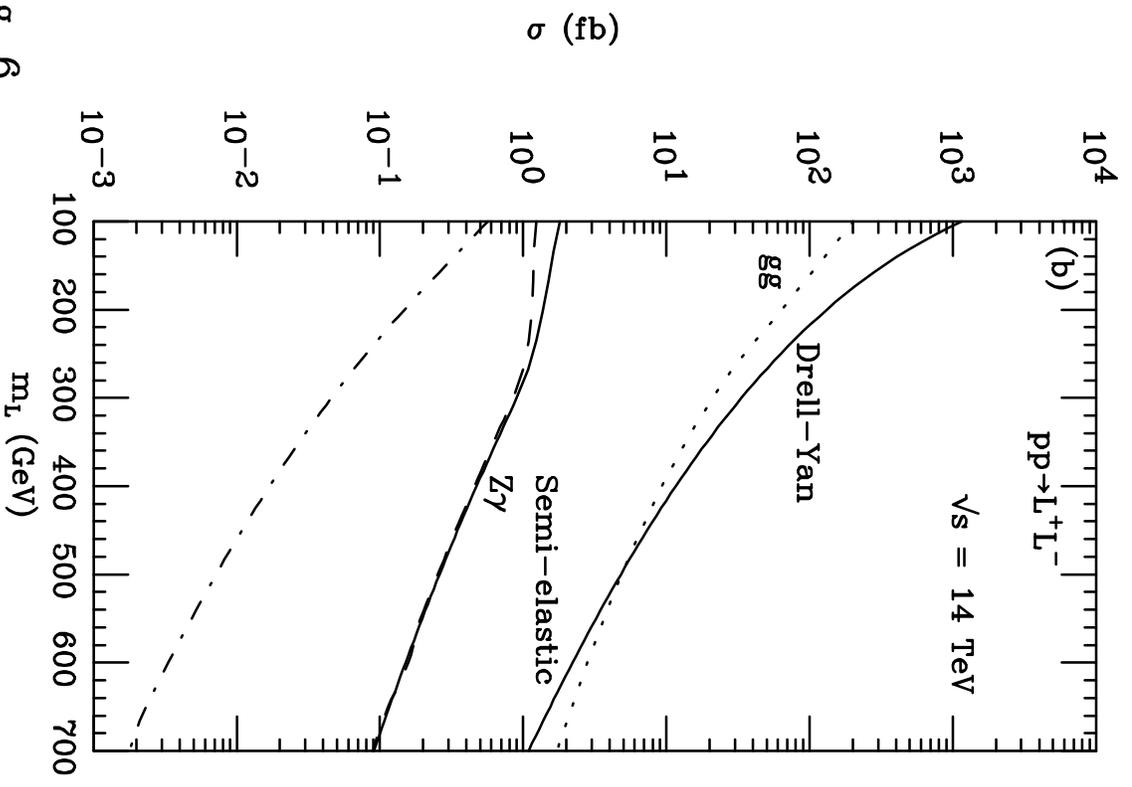
\begin{figure}
\caption[]{The total production cross section (in femtobarns) for a heavy
charged lepton pair in $pp$ collisions at the LHC ($\sqrt s = 14$ TeV) using
the non-leading Dawson distribution functions. In each of (a) and (b), the
upper
solid curve and the dotted curve show,  respectively, the Drell-Yan and gluon
fusion cross sections.
(a) The dashed and dot-dashed curves represent,
respectively, the longitudinal and transverse $Z$ boson  contributions
to the $Z \gamma$ fusion cross-sections for inelastic $pp$ collisions. The
lower solid curve shows the total inelastic $Z \gamma$ fusion cross-section.
(b) The curves have the same meaning as in (a) except that they are for
semi-elastic $pp$ collisions.
\label{fig6}}
\end{figure}


\begin{references}
\bibitem{eecol} For a review see {\sl e.g.} {\sl Proceedings of the Conference
 on Physics and Experiments at Linear Colliders}, Saariselk\"a, Finland,
 edited by R. Orava, Eerola and Nordberg, (World Scientific 1992); {\sl
 Proceedings of the IX th International Workshop on Photon-Photon Collisions},
 La Jolla, California, edited by D.O. Caldwell and H.P. Paar, World Scientific
 (1992); {\sl Proceedings of the Workshop on Physics and Experiments with
 Linear $e^+e^-$ Colliders}, Waikoloa, Hawaii, edited by F. A. Harris,
 S. L. Olsen, S. Pakvasa and X. Tata, World Scientific (1993).
\bibitem{gen} V. Barger, T. Han and J. Ohnemus, Phys. Rev. {\bf D37}, 1174
(1988) and references therein.
\bibitem{e6} J.L. Hewett and T.G. Rizzo, Phys. Rep. {\bf 183}, 193 (1989).
\bibitem{dy} S.D. Drell and T.M. Yan, Phys. Rev. Lett. {\bf25}, 316 (1970);
    Ann. Phys. (NY) {\bf66}, 578 (1971).
\bibitem{gf} S.S.D. Willenbrock and D.A. Dicus, Phys. Lett. {\bf B 156}, 429
    (1985).
\bibitem{gbf} S. Dawson and S.S.D. Willenbrock, Nucl. Phys. {\bf B 284}, 449
    (1987).
\bibitem{ww} E.J. Williams, Proc. R. Soc. London (A) {\bf139}, 163 (1933);
Phys. Rev. {\bf45}, 729(L) (1934); C.F. von Weizs\"{a}cker, Z. Phys. {\bf88},
    612 (1934).
\bibitem{elww} B.A. Kniehl, Phys. Lett. {\bf B 254}, 267 (1991).
\bibitem{us} G. Bhattacharya, P.A. Kalyniak and K.A. Peterson, {\sl Proc. of
the
   Workshop on Physics at Current Accelerators and Supercolliders},
   Argonne, Illinois, edited by J.L. Hewett, A.R. White and D. Zeppenfeld,
   (1993)p531; G. Bhattacharya, P.A. Kalyniak, and K.A. Peterson,
   {\sl Proceedings of the 16th
   Annual MRST (Montr\'eal-Rochester-Syracuse-Toronto) Meeting on High Energy
   Physics: What Next? Exploring the Future of High Energy Physics},
   Montr\'eal, Canada, edited by J.F. Cudell, K.R. Dienes, and B. Margolis
	(1994) p261.
\bibitem{do}M. Drees, R.M. Godbole, M. Nowakowski and S.D. Rindani, Phys. Rev.
 {\bf D50}, 2335 (1994); J. Ohnemus, T.F. Walsh and P.M. Zerwas, Phys. Lett.
 {\bf B328}, 369 (1994)
\bibitem{eboli}O.J.P. \'Eboli, G.C. Marques, S.F. Novaes, and A.A. Natale,
 Phys. Rev. {\bf D34}, 771 (1986)
\bibitem{kane}G.L. Kane, W.W. Repko, and W.B. Rolnick, Phys. Lett.
{\bf B148}, 367 (1984); J. Lindfors, Z. Phys. {\bf C28}, 427 (1985)
\bibitem{dawson}S. Dawson, Nucl. Phys. {\bf B 249}, 42 (1985)
\bibitem{brodsky}S.J. Brodsky, T. Kinoshita, and H. Terazawa, Phys. Rev.
{\bf D4}, 1532 (1971); H. Terazawa, Rev. Mod. Phys. {\bf 45}, 615 (1973)
\bibitem{ammr}G. Altarelli, G. Martinelli, B. Mele and R. R\"{u}ckl, Nucl.
     Phys. {\bf B262}, 204 (1985).
\bibitem{johnson}P.W. Johnson, F.I. Olness and W.K. Tung, Phys. Rev. {\bf D36},
 291 (1987)
\bibitem{godbole}R.M. Godbole and F.I. Olness, Int. J. Mod. Phys. {\bf A 2},
1025, (1987)
\bibitem{hmrs}P.N. Harriman, A.D. Martin, R.G. Roberts and W.J. Stirling,
Phys. Rev. {\bf D42}, 798 (1990).
\bibitem{dhkz}K. Hagiwara, S. Komamiya and D. Zeppenfeld, Z. Phys. {\bf C 29},
  115 (1985); M. Drees and D. Zeppenfeld, Phys. Rev. {\bf D 39}, 2536 (1989).
\bibitem{barn}R.M. Barnett {\sl et al.}, {\sl Proceedings of the Summer Study
on High Energy Physics in the 1990s}, Snowmass, Colorado (World Scientific
1988)
\bibitem{alfs}W. Marciano, Phys. Rev. {\bf D29}, 580 (1984).
\bibitem{effth}M. Chanowitz, M. Furman and I. Hinchliffe, Phys. Lett. {\bf
B78}, 285 (1978); Nucl. Phys. {\bf B153}, 402 (1979)
\bibitem{cahn}R.N. Cahn and S. Dawson, Phys. Lett. {\bf B136}, 196 (1984);
Erratum Phys. Lett. {\bf B138}, 464 (1984).
\bibitem{hinch}I. Hinchliffe, Int. J. Mod. Phys. {\bf A4}, 3867 (1989).
\end{references}
\end{document}